\documentclass[11pt]{article}
\usepackage{epsf,epsfig,psfig,float,amssymb,latexsym}
\usepackage{graphics,psfrag}
\newcommand{\La}{{\cal L}}

\newcommand{\cM}{{\cal M}}
\newcommand{\cH}{{\cal H}}
\newcommand{\cD}{{\cal D}}

\newcommand{\vp}{{\mathbf{p}}}
\newcommand{\vq}{{\mathbf{q}}}

\newcommand{\vQ}{{\mathbf{Q}}}

\newcommand{\vk}{{\mathbf{k}}}

\newcommand{\be}{\begin{equation}}
\newcommand{\ee}{\end{equation}}
\newcommand{\ba}{\begin{eqnarray}}
\newcommand{\ea}{\end{eqnarray}}

\newcommand{\nn}{\nonumber}

\newcommand{\vs}{\vspace{-0.20cm}}

\begin{document}

\thispagestyle{empty}

\vspace{2cm}

\begin{center}
{\Large{\bf Finite width effects in $\phi$ radiative decays}}
\end{center}
\vspace{.5cm}

\begin{center}
{\large Jos\'e A. Oller\footnote{email: oller@um.es}}
\end{center}

\begin{center}
{\it {\it Departamento de F\'{\i}sica. Universidad de Murcia.\\ E-30071,
Murcia. Spain.}}
\end{center}
\vspace{1cm}

\begin{abstract}
\noindent
{\small{ We calculate the decay rates $\phi\rightarrow \gamma
    \pi^0\pi^0$ and $\gamma \pi^0\eta$ in very  good agreement with  the
    recent accurate experimental data. We also point out the necessity of a 
 $\phi\gamma K^0\bar{K}^0$ contact vertex beyond the pure $K^+K^-$ loop
    model. The decay widths $\phi \rightarrow \gamma
    f_0(980)$ and $\phi \rightarrow \gamma a_0(980)$ are also calculated taking into
    account the finite widths of the scalar resonances $f_0(980)$ and
$a_0(980)$. The latter are shown to be essential in order to obtain meaningful
    results.The resulting decay rates to $\gamma f_0(980)$
    and $\gamma a_0(980)$ are in good agreement
    with the experimental ones without
invoking isospin breaking in the couplings of the $f_0(980)$ and $a_0(980)$
resonances to the
$K^+ K^-$ and $K^0 \bar{K}^0$ channels, at odds with recent proposals. The derived
formula for calculating these $\phi$ radiative decay widths can  be also applied
in their own experimental analyses in order to obtain more precise results.
}}
\end{abstract}

\vspace{2cm}


\newpage

\section{Introduction}
\def\theequation{\arabic{section}.\arabic{equation}}
\setcounter{equation}{0}

In this article we study the $\phi$ radiative decays to $\gamma$ and two
neutral pseudoscalars and the related rates to $\gamma f_0(980)$ and $\gamma
a_0(980)$. 
These decays were first observed experimentally
by the CMD-2 \cite{akh} and SND \cite{snd98,exppi0,expeta} collaborations at Novosibirsk
by measuring the invariant mass distributions
$d\Gamma(\phi \rightarrow \gamma \pi^0\pi^0)/dm_{\pi\pi}$ and
$d\Gamma(\phi \rightarrow \gamma\pi^0 \eta)/d m_{\pi\eta}$, respectively, where
we indicate by $m_{M^0 N^0}$ the invariant mass of a pair of neutral pseudoscalars
$M^0$ and $N^0$. On the other hand, high statistics results have been recently
reported by the KLOE collaboration at DA$\Phi$NE in
refs.\cite{dafnef0,dafnea0} for the rates $\phi\rightarrow \gamma \pi^0\pi^0$
and $\gamma \pi^0\eta$, respectively.  These decays offer an additional
source of experimental information
on the non-trivial and so important low lying scalar mesons with vacuum quantum numbers
$0^{++}$. Our theoretical study generalizes the one undertaken in
ref.\cite{plb} about the $\phi\rightarrow \gamma K^0\bar{K}^0$ decay  in order
to reproduce the precise data of refs.\cite{dafnef0,dafnea0} on
$\phi\rightarrow \gamma\pi^0\pi^0$ and $\gamma \pi^0\eta$ and connects the previous
processes with chiral symmetry and unitarity. Ref.\cite{plb} was also followed
closely by ref.\cite{marco} where the aforementioned radiative decays to the
final states $\gamma \pi^0\pi^0$ and $\gamma \pi^0\eta$  were also
considered. 
For recent accounts on the application
of similar techniques to the scalar meson-meson and meson-baryon dynamics we refer the
reader to refs.\cite{anke,kyoto,jamin} where many references are presented and discussed.
Other studies of $\phi\rightarrow \gamma M^0 N^0$ decays are
refs.\cite{acha,lucio,truong,close,bra1,bra3,markushin,referee,gok,fazio,escri}.

In order to interpret the experimental results from refs.\cite{exppi0,expeta}
on the $\phi$ decays to
$\gamma f_0(980)$ and $\gamma a_0(980)$, ref.\cite{close2} has pointed out recently
the necessity of considering important isospin breaking corrections in the couplings
of the $f_0(980)$ and $a_0(980)$ resonances to the $K^+K^-$ and $K^0
\bar{K}^0$ channels. That is, $|g_{K^+K^-}^R|$ and $|g_{K^0\bar{K}^0}^R|$,
where $R$ represents either the $f_0(980)$ or $a_0(980)$ resonances, are
no longer equal as required by isospin symmetry but can be actually quite
different according to ref.\cite{close2}. Special emphasis is given to the
rather large decay width $\phi
\rightarrow \gamma f_0(980)$ together with the deviation from one, by around a factor four,
of the ratio $Br(\phi\rightarrow \gamma f_0)/Br(\phi\rightarrow \gamma a_0)$
as experimentally
established in refs.\cite{akh,exppi0,expeta,dafnef0,dafnea0}. 
Nevertheless, in ref.\cite{acha2} it is
clearly shown that possible isospin breaking effects in the quotient
$\Gamma(\phi \rightarrow \gamma \pi^0\pi^0))/\Gamma (\phi \rightarrow \gamma \pi^0\eta)$
cancel to a large extend.
As a result, this ratio is sensitive mostly to the isospin
eigenstates $a_0(I=1)$ and $f_0(I=0)$. Similar conclusions are obtained within the 
vector meson dominance model of ref.\cite{joe} as well. 
Other references treating the $f_0(980)$ and $a_0(980)$ mixing are \cite{speth,isos1,isos2,isos3}.

We show that one does not need to abandon
isospin symmetry in the couplings of the scalar resonances $f_0(980)$ and $a_0(980)$
to the $K\bar{K}$ channels so as to account for the experimental results on the
$\phi \rightarrow \gamma R$ decay widths. We emphasize the fact that the
threshold of $\gamma R$ is so close
to the mass of the $\phi(1020)$ resonance, that meaningful results for the
width $\Gamma(\phi\rightarrow \gamma R)$ can only result once the finite widths of the
$f_0(980)$ and $a_0(980)$ resonances are taken into account. It is important
to stress that the  decay width $\Gamma(\phi \rightarrow \gamma R)$ depends
cubically on the small photon momenta and hence small changes in the nominal masses
of the $f_0(980)$ or $a_0(980)$ resonances, e.g. as compared to their widths or to the
difference between the real parts of the pole positions and the peaks of the
scattering amplitudes, imply dramatic
variations on the resulting $\phi$ radiative decay widths to $\gamma R$ if the standard
two body decay formula were used.
In sec.\ref{sec:direct} we calculate the $\phi$ decays to
 $\gamma \pi^0\eta$, $\gamma \pi^0\pi^0$ and $\gamma K^0\bar{K}^0$ where we 
connect with and extend the results already given in
 refs.\cite{plb,marco}. We also establish a new contribution beyond the pure $K^+
 K^-$ loop model of ref.\cite{acha} due to a contact $\phi\gamma K^0\bar{K}^0$
 vertex.  In sec.\ref{sec:gf}, making use of the results of the previous
 section, we take into account the finite
width effects of the $f_0(980)$ and $a_0(980)$ resonances
on the decays $\phi \rightarrow \gamma R$ and derive the corresponding formulae.
We also consider  the connection of the obtained
formula for  $\Gamma(\phi\rightarrow \gamma R)$ to the invariant
mass distributions $d\Gamma(\phi \rightarrow \gamma M^0 N^0)/dm_{M^0 N^0}$. 
Finally,  our results are discussed in sec.\ref{sec:wfr} and the conclusions are given in
sec.\ref{sec:conc}.


\section{ $\Gamma(\phi \rightarrow \gamma M^0 N^0)$ rates}
\label{sec:direct}

\begin{figure}[htb]
\centerline{\epsfig{file=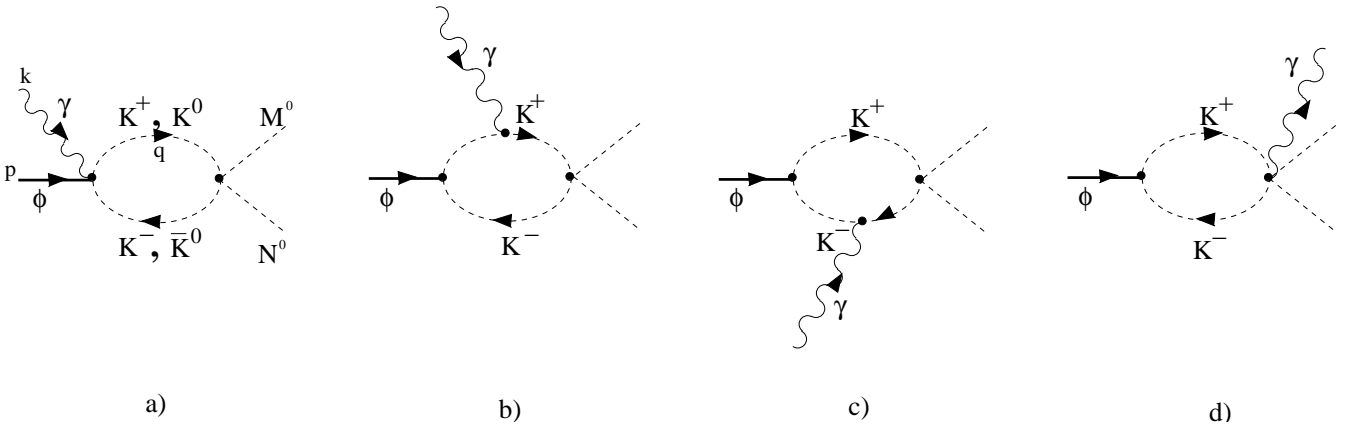,width=7.0in}}
\vspace{.3cm}
\caption[pilf]{\protect \small
  Set of diagrams for the calculation of $\phi \rightarrow \gamma M^0 N^0$ through kaon loops. 
\label{fig:loop}}
\end{figure}

We are interested in studying the $\phi\rightarrow\gamma
M^0N^0$ decays. The final state interactions (FSI) of the
two pseudoscalar mesons are assumed to be dominated by their strong S-wave
amplitudes which contain the $f_0(980)$ and $a_0(980)$ scalar
resonances. Then gauge invariance requires, as noted in
refs.\cite{acha,acha3,close} with respect to the related decays
$\phi\rightarrow \gamma R$ as well, that the transition amplitude 
$\phi(1020)\rightarrow \gamma M^0 N^0$ must have the form \cite{plb,marco}:
\be
\label{gaugein}
{\cal M}\left[\phi(p) \rightarrow \gamma(k) M^0N^0 (Q)\right]=
H(Q^2)\left[g^{\alpha\beta} (p\cdot k)-p^\alpha k^\beta\right]
\epsilon^{\lambda}(\gamma)_\alpha \epsilon^{\nu}(\phi)_\beta~.
\ee   
 The kinematics is indicated in fig.\ref{fig:loop} and the four-momentum $Q$ 
corresponds to the total one of the pair of pseudoscalars $M^0N^0$. 

As in refs.\cite{acha,bra1,plb} we consider the $K^+ K^-$ meson loop contribution
to the  decay $\phi \rightarrow \gamma M^0 N^0$, fig.\ref{fig:loop}, 
since the $K^+K^-$ system strongly couples to both the $\phi$ and $R$
resonances. In addition, since the $K^0 \bar{K}^0$ couples with the same
strength as the $K^+K^-$
state both to the scalar and  $\phi$ resonances, we  also allow for
a neutral kaon loop  contribution, as indicated in fig.\ref{fig:loop}a, where
the vertex on the left of the figure is discussed in length below.

To simplify the discussions, we will consider the isospin channels 
separately, since they do not mix by final state interactions in the isospin
limit,  so that $H^I(Q^2)$ is the invariant
function in the same isospin channel as the corresponding scalar resonance $R$ from the 
kaon loops of fig.\ref{fig:loop}. We now calculate the function ${\cal H}^I(Q^2)$ 
defined such that $H^I(Q^2)={\cal H}^I(Q^2) t^R_{K^+K^- \rightarrow M^0
  N^0}$,  with $t^R_{K^+K^- \rightarrow M^0
  N^0}$  the isospin amplitude 
$-t^I_{K\bar{K}\rightarrow M N}\,C^I_{M^0N^0}/\sqrt{2}$ where the
$K\bar{K}$ and $MN$ states are in 
the isospin channel $I$ of the scalar resonance $R$, namely, $I=0$ for the
$f_0(980)$ and $I=1$ for the $a_0(980)$, and $-1/\sqrt{2}$ and $C^I_{M^0N^0}$
are Clebsch-Gordan coefficients. For the $\gamma \pi^0\eta$ final
state only the $I=1$ channel takes place and the $I=0$ one  is the only contribution to 
$\gamma \pi^0\pi^0$. For the $\gamma K^0\bar{K}^0$ final state both contributions sum
up. 

In ref.\cite{plb} we took as interacting Lagrangian, 

\be
\label{lag}
\La=2 e g_\phi A^\mu \phi_\mu K^+ K^- 
-i\,(e A_\mu+g_\phi \phi_\mu) (K^-\partial^\mu K^+ - \partial^\mu K^- K^+)~,
\ee 
upon making the $\phi$ and $K^+ K^-$ interactions gauge invariant. 

We now consider the more refined approach of ref.\cite{conga}, where the 
couplings of the octet of pseudoscalars to the octet of vector resonances are
given at order $p^2$ by taking into account chiral symmetry  in a chiral power 
expansion. The couplings of the singlet vector resonance $\omega_1$ cancel 
at this order exactly. Making use of ideal  mixing we can write the $\phi$ resonance 
as:
\be
\label{mixing}
\phi=-\frac{2}{\sqrt{6}}\omega_8+\frac{1}{\sqrt{3}}w_1~,
\ee 
and thus the couplings of the $\phi$ resonance from ref.\cite{conga} are 
the ones of the $\omega_8$ resonance times $-2/\sqrt{6}$.  The 
tree level matrix element for the contact vertex $\phi \gamma K^+K^-$, 
corresponding to the left vertex of fig.\ref{fig:loop}a, 
$M(\phi\rightarrow \gamma K^+ K^-)_{contact}$, calculated from the formalism
of ref.\cite{conga} is:
\be
\label{phiconga}
M(\phi\rightarrow \gamma K^+ K^-)_{contact}=
-\frac{\sqrt{2}\,e M_\phi G_V}{f^2}\epsilon(\gamma)\cdot
\epsilon(\phi) -
\frac{\sqrt{2}\,e}{f^2 M_\phi}\left(\frac{F_V}{2}-G_V\right)
p_\alpha \epsilon(\phi)_\beta\left(k^\alpha \epsilon(\gamma)^\beta
-k^\beta \epsilon(\gamma)^\alpha\right)~,
\ee
where $f\simeq 93$ MeV is the weak pion decay constant and $G_V$ measures the
strength of the transition  $\phi\rightarrow K^+ K^-$.  The experimental width 
$\Gamma(\phi \rightarrow K^+K^-)$ is reproduced with $G_V=55$ MeV. Let us note
that the second structure in the equation above is gauge invariant by itself. On the
other hand, vector meson dominance requires $F_V=2 G_V$ \cite{conga} and in
this limit the coefficient in front of the second term of eq.(\ref{phiconga})
is zero. At the chiral order considered in ref.\cite{conga}, there is no direct
coupling $\phi\gamma K^0\bar{K}^0$ although at higher orders this is no longer
the case and can mimic e.g. $\phi\gamma K^0\bar{K}^0$ vertices generated
through the exchange of vector resonances, see e.g.\cite{bra1}.  

We consider first the $K^+K^-$ loops of fig.\ref{fig:loop} that
are originated from the surviving local term of eq.(\ref{phiconga}),
fig.\ref{fig:loop}a, plus the
Bremsstrahlung ones, diagrams \ref{fig:loop}b and c, and fig.\ref{fig:loop}d. 
We follow the treatment
of ref.\cite{plb} where it is shown that the T-matrix element
$t^R_{K^+K^-\rightarrow K^+K^-}$ of ref.\cite{npa} factorizes on-shell in these diagrams. The
reason is, as proved in ref.\cite{plb}, that the off-shell parts of the
T-matrices of ref.\cite{npa} do not contribute to the
coefficient of the term in eq.(\ref{gaugein}) proportional to $p^\alpha
k^\beta$, which is the function $H(Q^2)$ as required by gauge
invariance.\footnote{It is shown in
  ref.\cite{nd} that the contributions from crossed channels exchanges of
  pairs of pseudoscalars and resonances tend to cancel in the S-waves with
  $I=0,$ 1 and $1/2$ below roughly 1 GeV.}

Motivated by the self-invariant structure of the second term of
eq.(\ref{phiconga}), we consider for each isospin channel a general contact 
$\phi\gamma K\bar{K}$ interaction with the same structure but parameterized by a
coupling $\zeta_I$, 
\ba
\label{local}
V_I(\phi\gamma K\bar{K})=-
\frac{\sqrt{2}e\,\zeta_I}{f^2 M_\phi} p_\alpha \epsilon(\phi)_\beta 
\left(k^\alpha \epsilon(\gamma)^\beta-k^\beta \epsilon(\gamma)^\alpha\right)~.
\ea
The dependence of $\zeta_I$ on the isospin channel implies a possible non-vanishing
$\phi \gamma K^0\bar{K}^0$ local term since the Clebsch-Gordan coefficient for
$K^0\bar{K}^0$ changes sign from $I=0$ to $I=1$ while that of $K^+K^-$ is the
same for both isospins. Similar structures for $\gamma\pi^0\pi^0$ and
$\gamma\pi^0\eta$ are not considered since they are expected to be suppressed
by the OZI rule.

\begin{figure}[htb]
\centerline{\epsfig{file=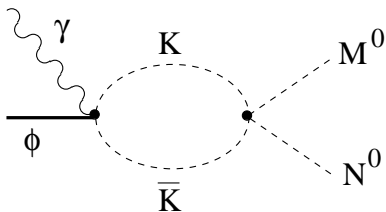,width=2.0in}}
\vspace{.3cm}
\caption[pilf]{\protect \small
  Diagram for $\phi \rightarrow \gamma M^0 N^0$ with the contribution of 
eq.(\ref{local}), which is gauge invariant by itself, in the vertex on the left.
This contribution was considered in ref.\cite{marco} to some extend but not in 
refs.\cite{plb,lucio,truong,close}.  Note that in the
vertex on the right the full amplitude $t^R_{K^+K^- \rightarrow M^0 N^0}$ is
considered.   
\label{fig:loop2}}
\end{figure}

We are now in position to evaluate  the loop of fig.\ref{fig:loop2}
with the contact term $V_I(\phi\gamma K \bar{K})$ of eq.(\ref{local}), 
in the vertex on the left of the diagram. The strong amplitude 
$t^R_{K^+  K^- \rightarrow M^0  N^0}$ of ref.\cite{npa} 
($t^R_{K^0\bar{K}^0 \rightarrow M^0 N^0}$ is
just proportional to the former) factorizes on-shell since
the off-shell part, as in the case of the pure strong interacting problem
treated in detail in ref.\cite{npa}, just renormalizes the coupling on the left
part of the loop, namely, the $\zeta_I$ factor of 
eq.(\ref{local}). This is due to the fact that it only gives rise to tadpole
like diagrams. Since the $\zeta_I$ factor is a free one this renormalization
process is not relevant for our present purposes and is just reabsorbed in
the final value of $\zeta_I$. Since both vertices on the diagram of 
fig.\ref{fig:loop2} factorize 
with their physical renormalized values, the one on the left does not involve 
any $K\bar{K}$ momentum, see eq.(\ref{local}), we are then
left with the logarithmic divergent loop integral, 
\be
\label{kloop}
i\int \frac{d^4 q}{(2\pi)^4}\frac{1}{\left(q^2 -m_{K^+}^2+i0^+\right)
\left( (Q-q)^2-m_K^2+i0^+\right)}~.
\ee
 In ref.\cite{marco} this integral 
was taken the same as the one fixed in ref.\cite{npa} for the strong
interactions.  Nevertheless, one should point out that unitarity and analyticity, 
with the latter restricted to the presence of the unitarity or right hand cut,
allow the presence of an extra subtraction constant, $\delta G^I$, in the integral of
eq.(\ref{kloop}) as compared to $G_{K \bar{K}}$ of ref.\cite{npa}. We show now this point. 

Since the vertex of eq.(\ref{local}) is gauge invariant by itself, we
consider its contributions separately. Thus, it is characterized by the
corresponding invariant function $\widetilde{H}^I$, a contribution to the total
one $H^I$. Let us denote by $\widetilde{H}^I_\alpha$ the invariant function
corresponding to the $\phi \rightarrow \gamma \alpha$ decay where $\alpha$
indicates any channel that can couple strongly to the $K \bar{K}$ channel and 
with the same well defined isospin. We treat these
invariant functions as form factors 
depending only on $Q^2$ since this is the
dependence obtained from unitarity loops as those of fig.\ref{fig:loop2} that we 
want to analyze.\footnote{Note that
  the vertex on the right is the full strong T-matrix with its own unitarity
  cuts.} Then unitarity requires:
\be
\hbox{Imag}\left[\widetilde{H}^I_\alpha\right]=\sum_{\beta}(\widetilde{H}^I_\beta)^*\,
\rho_\beta \,t_{\beta\rightarrow \alpha}\, \theta(Q^2-W_\beta^2)~,
\ee 
where $\rho_\beta$ is the phase space factor of channel $\beta$, 
$\rho_\beta=q_\beta/(8\pi\sqrt{Q^2})$. In ref.\cite{jpsi,palomar} it is shown
that when the T-matrix appearing on the
right side of the previous equation has no crossed cuts due to crossed channels, as the
one of refs.\cite{npa,iamprd}, $\widetilde{H}^I_\alpha$, collected in the vector
column $\widetilde{H}^I$, can be represented as:
\be
\label{jpsi}
\widetilde{H}^I=\left[1+K^I(Q^2)\cdot G(Q^2)\right]^{-1} R_I(Q^2)~,
\ee 
where it is important to remark that $R_I(Q^2)$ is a vector of functions without any cut. 
In addition, $T^I(Q^2)=\left[1+K^I(Q^2)G(Q^2)\right]^{-1}\cdot K^I(Q^2)$ with $T^I(Q^2)$ the
strong T-marix for the S-wave meson-meson scattering with isospin $I$, 0
or 1. In ref.\cite{npa} $K^I(Q^2)$ corresponds to the lowest order CHPT
meson-meson amplitudes \cite{wein,gl} while in ref.\cite{iamprd} together with
this contribution one has the local terms calculated from ${\cal O}(p^4)$ CHPT,
see refs.\cite{npa,iamprd} for explicit formulae.
In addition $G(Q^2)$ is a diagonal matrix of unitarity loops such
that for $I=0$, $G_{11}(Q^2)$ is the loop analogous to  eq.(\ref{kloop}) but for
$\pi\pi$($\alpha=1$) and $G_{22}(Q^2)$ corresponds to the $K\bar{K}$($\alpha=2$) channel. 
Analogously for $I=1$, $G_{11}(Q^2)$ corresponds to
$\pi\eta$($\alpha=1$) and $G_{22}(Q^2)$
to $K\bar{K}$($\alpha=2$) and is equal to $G_{22}(Q^2)$ with $I=0$.

On the other hand, because of the aforementioned factorization of the T-matrix of 
ref.\cite{npa},  $\widetilde{H}^I_\alpha$ can be expressed as:
\be
\label{h'}
\widetilde{H}^I=\lambda_I-T_I \, {\cal G}^I(Q^2)\,\lambda_I ~,
\ee 
with $\lambda_I$ the vector column of direct couplings to $\phi \gamma K\bar{K}$ after 
removing the tensor structures like in the definition of $\widetilde{H}^I$,
 that is $(\lambda_I)^T \propto (0,\zeta_I)$.  
 On the other hand, ${\cal G}^I_{ii}(Q^2)$ can differ at most from
$G_{ii}(Q^2)$ in a subtraction constant which we express by writing ${\cal
  G}^I=G+\delta G^I$. Notice that because $(\lambda_I)_1=0$ only the matrix
element of $\delta G^I$
for the $K\bar{K}$ channel is relevant, the only one that we will keep in the
following.  Now, taking into account the explicit 
expression of $T^I$ in terms of $G$ and $K^I$, one has:
\ba
\label{ayay}
\widetilde{H}^I&=&\left(1-\left[1+K^I G\right]^{-1}\left\{K^I G+K^I\delta
    G^I\right\}\right)
\lambda_I=
\left[1+K^IG\right]^{-1}\lambda_I-\left[1+K^I G\right]^{-1}K^I\delta G^I
    \lambda_I \nn\\
&=&\left[1+K^I G\right]^{-1}(1-K^I \delta G^I)\lambda_I ~,
\ea
and hence $R_I=(1-K^I \delta G^I)\lambda_I$. Let us note that $R_I$ has no cuts, as it should,
 since neither $K^I$, $\delta G^I$ or $\lambda_I$ have any cut.

As a result of this digression, we will keep $\delta G^I$ as free parameters in addition to
$\zeta_I$. This amounts to four free parameters, two for each isospin
channel, although in the end we will show that two of them will turn out to be fixed. 
Although eq.(\ref{h'}) is derived by taking into account the off-shell form
of the T-matrices of ref.\cite{npa}, we will
also use it for the more involved T-matrices of
refs.\cite{iamprd} since the equivalent form given in eq.(\ref{ayay}) is
equally valid for all of them. That is, the column matrix $R_I(Q^2)$ is fixed 
to be $(1-K^I \delta G^I)\lambda_I$, eq.(\ref{ayay}), for both the T-matrices of
ref.\cite{npa} and ref.\cite{iamprd}. While for the former strong amplitudes this expression
for $R_I$ is a strict
derivation due to its specific off-shell parts that enter in the unitarity
loops, for the latter ones $R_I$ is fixed  by analogy.

Summing the $K^+K^-$ loop contribution of figs.\ref{fig:loop}a, b, c and d plus
the $K\bar{K}$ loop of fig.\ref{fig:loop2}, $\widetilde{H}^I$, we can then
write:
\be
\label{totalH}
H^I(Q^2)=\left\{\frac{\sqrt{2}\, eM_\phi G_V}{4\pi^2 f^2 m_{K^+}^2}I(a,b)-
\frac{2\,e\, \zeta_I }{f^2 M_\phi} {\cal G}^I_{K\bar{K}}(Q^2)\right\}t^R_{K^+K^-
\rightarrow M^0 N^0}~,
\ee
where for the $\gamma \pi^0\pi^0\,(\pi^0\eta)$ final state only $I=0\, (1)$
contributes while for $\gamma K^0\bar{K^0}$ one has to sum both $I=0$ and 1
contributions together. From this expression it follows that 
${\cal H}^I(Q^2)=H^I(Q^2)/t^R_{K^+K^-
\rightarrow M^0 N^0}$ is given by:
\ba
\label{figamaR}
\cH^I(Q^2)&=&\frac{\sqrt{2}\, e M_\phi G_V}{4\pi^2 f^2 m_{K^+}^2}I(a,b)-
\frac{2\,e\,\zeta_I}{f^2 M_\phi} {\cal G}^I_{K\bar{K}}(Q^2)~.
\ea
Including the three-body phase space and taking into account
eq.(\ref{totalH}), 
the width $\Gamma(\phi\rightarrow \gamma M^0 N^0)$ for $M^0 N^0=\pi^0\pi^0$
or $\pi^0\eta$ can be expressed as:
\be
\label{gammam0n0}
\Gamma(\phi\rightarrow \gamma M^0 N^0)\!=\!\int \!\!d\sqrt{Q^2}\,\,
\frac{\alpha |\vk|^3 |\vp_M|}{6\pi^2f^4}  
\left|\sum_R \left(\frac{M_\phi G_V}{4\pi^2
    m_{K^+}^2}I(a,b)-\frac{\sqrt{2}\,\zeta_I }{M_\phi}{\cal G}^I_{K\bar{K}}(Q^2)
\right)\, t^R_{K^+K^-\rightarrow M^0N^0}\right|^2~,
\ee
where $|\vp_M|$ is the momentum of the $M^0 N^0$ system in their center of
mass frame and  $\alpha$ is the electromagnetic fine structure constant. 
 Possible symmetric identity factors 1/2 are omitted in
eq.(\ref{gammam0n0}) since the transition matrix element
$|t^R_{K^+K^-\rightarrow M^0  N^0}|^2$ is taken from refs.\cite{npa,iamprd}
and such factors are already included in its normalization. In the previous 
expression the sum is restricted over those isospin channels included in $M^0N^0$. For the 
$\gamma K^0\bar{K}^0$ final state case, $M^0N^0=  K^0\bar{K}^0$, one has
to add to the term between bars in 
eq.(\ref{gammam0n0}) the 
contribution $+(\zeta_0-\zeta_1)/\sqrt{2} M_\phi$ from the direct local terms of 
eq.(\ref{local}) with $I=0$ and 1.

\section{$\Gamma(\phi \rightarrow \gamma R)$ decay widths}
\label{sec:gf}
\def\theequation{\arabic{section}.\arabic{equation}}
\setcounter{equation}{0}

As in the previous section, see also refs.\cite{acha,acha3,close,plb}, gauge
invariance requires that
the transition amplitude $\phi(1020)\rightarrow \gamma R$ must have the form:
\be
\label{gaugeinR}
{\cal M}\left[\phi(p) \rightarrow \gamma(k) R (Q)\right]=
H^R(Q^2)\left[g^{\alpha\beta} (p\cdot k)-p^\alpha k^\beta\right]
\epsilon^{\lambda}(\gamma)_\alpha \epsilon^{\nu}(\phi)_\beta~.
\ee
where now $Q$ is the total four-momentum of the resonance $R$. The function
$H^R(Q^2)$ can be obtained from the expression given above for
 $H^I(Q^2)$ replacing $t^R_{K^+K^-\rightarrow M^0N^0}$ by the
coupling of the $K^+K^-$ channel to the $R$ resonance, $g^R_{K^+K^-}$. That
is, we are extracting the resonance pole contribution from $t^R_{K^+K^-\rightarrow
  M^0N^0}$. Although not explicitly shown $g^R_{K^+K^-}$ can depend on the
energy, $Q^2$, flowing to the resonance $R$ which is not fixed since its mass 
 is not well defined due to the finite widths of $R$. Indeed, the latter 
 gives rise to important effects due to  the proximity of the mass of the
$\phi(1020)$ resonance to the nominal ones of the $f_0(980)$ and
$a_0(980)$. Let us consider this in detail.

Since the coupling
 $g^R_{K^+K^-}$ appears as a factor in $H^R(Q^2)$, then we have that  
 $H^R(Q^2)=\cH^I(Q^2)\,g_{K^+K^-}^R$. It is also  convenient to define the quantity $\cM'$ 
 so that $\cM(\phi\rightarrow \gamma R)=\cM'\,g_{K^+K^-}^R$. Then, 
the width $\Gamma(\phi \rightarrow\gamma R)$ can be expressed as:
\be
\label{width}
\Gamma(\phi \rightarrow \gamma R)=\frac{1}{3}\sum_{\nu=1}^3 \sum_{\lambda=1}^2
\int \frac{d \vk}{(2\pi)^3 2 |\vk|}\int \frac{dQ^0}{2 Q^0} f_R(Q^0) 
\frac{|\cM' g_{K^+ K^-}^R|^2}{2 M_\phi} (2\pi)\delta(M_\phi-|\vk|-Q^0)~.
\ee
Of course, because of three-momentum conservation, $\vQ=-\vk$ in the center 
of mass frame where the $\phi$ is at rest. We have included energy 
distributions for the scalar resonances $f_R(Q^0)$ in order 
to take care of their finite widths. This is not considered
in the literature for the calculation $\Gamma(\phi \rightarrow \gamma R)$ since well
defined values for the masses of the scalar resonances $f_0(980)$ and $a_0(980)$  are
assumed and the standard two body decay formula, eq.(\ref{gama2body}), is
then applied, see e.g. \cite{close,close2,joe}. We should note that
this is a very unstable procedure since the decay width scales as $|\vk|^3$, 
 where $|\vk|$ is the photon three-momentum:
\be
\label{k}
|\vk|=\frac{M_\phi^2 - Q^2}{2 M_\phi}~.
\ee
Allowing a change of $m_R$ of just 10 MeV from the central value 
$m_R=980$ MeV, the 
quantity $|\vk|$ changes by a factor of 1.7 and $|\vk|^3$ by a
factor of five. The standard two body decay formula can be obtained 
from the more general eq.(\ref{width}) by taking 
$f_R(Q^0)=\delta(Q^0-\sqrt{\vQ^2+m_R^2})$.

\begin{figure}[htb]
\psfrag{f0f0}{\hspace{-0.5cm}$\pi f_{f_0}(Q^0)/2Q^0$}
\psfrag{f0a0}{\hspace{-0.5cm}$\pi f_{a_0}(Q^0)/2Q^0$}
\psfrag{E(MeV)}{\begin{tabular}{l} \\ \hspace{-0.5cm}{\small $Q^0$ (MeV)}\end{tabular}}
\centerline{\epsfig{file=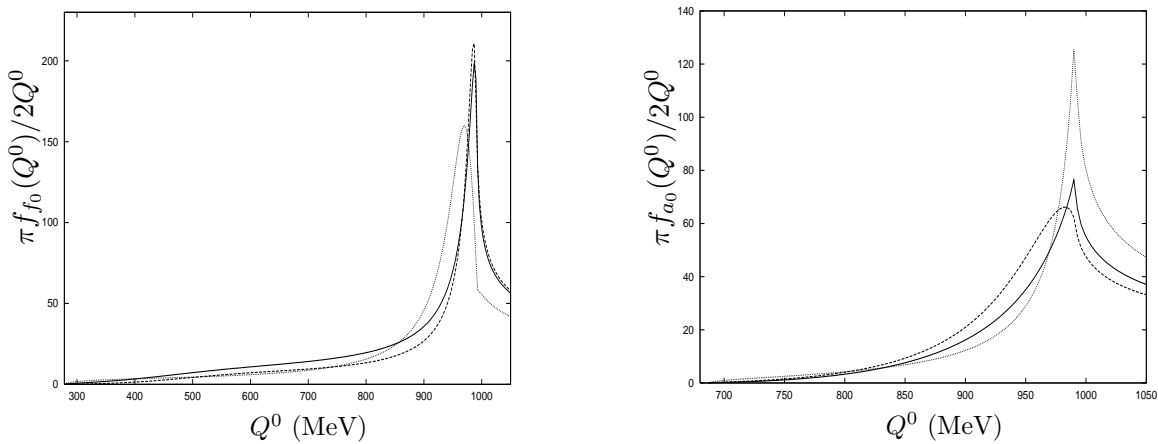,width=6.0in}}
\vspace{0.2cm}
\caption[pilf]{\protect \small
 $\pi f_R(Q^0)/2 Q^0={\hbox{Imag}}[t^R_{K^+ K^- \rightarrow K^+ K^-}]$ for the
 $f_0(980)$, 
left panel, and $a_0(980)$, right panel. 
 The solid and dashed lines are calculated from the T-matrices
 of refs.\cite{iamprd,npa}, respectively, and the thin dotted  ones 
from the resonant form given in eq.(\ref{achaprop}) following ref.\cite{acha} and the 
 experimental analyses \cite{exppi0,expeta}.  The values of the masses and
 couplings, necessary to apply eq.(\ref{achaprop}), are determined by the best fits
 of the latter references.
\label{fig:f_0}}
\end{figure}

In order to fix $f_R(Q^0)$ we invoke unitarity which implies:
\ba
\label{uni0}
2\,\hbox{Imag}[t_{K^+ K^-\rightarrow K^+K^-}^R]=\langle K^+ K^- | T^\dagger \, T | 
K^+ K^- \rangle~.
\ea
Now, assuming that unitarity is saturated by the exchange of the corresponding scalar 
resonance $R$, it follows that,
\ba
\label{uni1}
2\,\hbox{Imag}[t_{K^+ K^-\rightarrow K^+K^-}^R]&=&\int \frac{d^4 q}{(2\pi)^3 2 q^0}
f_R(q^0) (2\pi)^4 \delta^4(Q-q)\,|g_{K^+ K^-}^R|^2\nn\\
&=&\frac{\pi}{Q^0}f_R(Q^0)\,|g_{K^+ K^-}^R|^2~,
\ea
and then,
\ba
\label{uni2}
\frac{1}{2Q^0} f_R(Q^0)\,|g_{K^+ K^-}^R|^2=\frac{1}{\pi}\hbox{Imag}
[t_{K^+ K^- \rightarrow K^+ K^-}^R(Q^2)]~.
\ea
 The previous equation gives our final form for $f_R(Q^0)$. We show 
in fig.\ref{fig:f_0} the combination 
$\pi f_R(Q^0)/2 Q^0=\hbox{Imag}[t_{K^+ K^- \rightarrow K^+ K^-}^R(Q^2)]$ for the
$f_0(980)$, left panel, and $a_0(980)$, right panel. The dashed lines correspond to the 
matrix elements $t_{K^+ K^- \rightarrow K^+ K^-}^R(Q^2)$ given in ref.\cite{npa}, where 
the only one free parameter is fixed in this reference by a successful study of
the strong $\pi\pi$, $K\bar{K}$ and $\pi\eta$ S-wave meson-meson interactions.  These T-matrices have
 been used by now in a whole series of studies about the scalar sector 
stressing the role of 
unitarity and chiral symmetry in order to understand the scalar dynamics.  
These studies comprise strong interactions \cite{npa}, fusion of two 
photons to $\pi^0\pi^0$, $\pi^+ \pi^-$, $K^+ K^-$, $K^0\bar{K}^0$ and $\pi^0\eta$ \cite{gamma}, 
$\phi$ radiative 
decays \cite{plb,marco}, $J/\Psi$ decays \cite{jpsi} and $pp$ collisions 
\cite{anke}. In the same figure we also show by the solid lines, 
$\hbox{Imag}[t_{K^+ K^- \rightarrow K^+ K^-}^R(Q^2)]$
from the T-matrices of ref.\cite{iamprd} with the use of the Inverse Amplitude 
Method \cite{iam}. We also compare our choice for $\pi f_R(Q^0)/2 Q^0$ from the T-matrices of 
refs.\cite{npa,iamprd}, with the one that results by taking,
\ba
\label{achaprop}
 \hbox{Imag}[t_{K^+ K^- \rightarrow K^+ K^-}^R(Q^2)]&=&-\hbox{Imag}\left[
 \frac{(g^R_{K^+K^-})^2}{\cD_R(Q^2)}\right]~, \nn \\
 \cD_R(Q^2)&=&Q^2-m_R^2-\hbox{Re}\Pi(m_R^2)+\Pi(Q^2)~,
\ea
from ref.\cite{acha}, since this resonant form was followed 
in the experimental references \cite{akh,snd98,exppi0,expeta,dafnef0,dafnea0} to fit their data. 
The energy distribution resulting from eq.(\ref{achaprop}) is shown in 
fig.\ref{fig:f_0} by the thin dotted lines. The free parameters in eqs.(\ref{achaprop}), 
two masses and four couplings, are fixed to the central values given by the best fits of 
refs.\cite{exppi0,expeta} 
to $d\Gamma(\phi\rightarrow \gamma \pi^0\pi^0)/dm_{\pi \pi}$ and 
$d\Gamma(\phi\rightarrow \gamma \pi^0\eta)/dm_{\pi \eta}$, respectively. The $\Pi_R(Q^2)$
is the sum of two-meson self energies, $\pi\pi$ and $K\bar{K}$ for $I=0$ and 
$\pi\eta$ and $K\bar{K}$ for $I=1$ as in refs.\cite{exppi0,expeta}. Explicit formulae
for $\Pi(Q^2)$ can be found in ref.\cite{acha}. It is interesting to remark the presence
of a long tail
in each of the energy distributions $f_R(Q^0)$ towards low invariant masses.
This is a result of keeping the real parts of the unitarity two-meson loops 
\cite{npa,iamprd} or
the real parts in the two-meson self energies in $\cD_R(Q^2)$, eq.(\ref{achaprop})
\cite{acha}.

 Then, 
 $\Gamma(\phi \rightarrow \gamma R)$, eq.(\ref{width}), can be written 
 from eqs.(\ref{gaugeinR}) and (\ref{uni2}) as:
\ba
\label{widthR}
\Gamma(\phi \rightarrow \gamma R)&=&
\int \frac{|\vk| d|\vk|}{24\pi^2M_\phi}\hbox{Imag}\,[t_{22}^R(Q^2)] 
\left|\cH^I(Q^2)\right|^2(M_\phi^2-Q^2)^2~,
\ea
where we have used the shorter notation $t_{22}^R$ instead of $t_{K^+ K^- \rightarrow 
K^+ K^-}^R$. 
Of course, one should keep in mind that $Q^2$ and $|\vk|$ are not independent 
variables but related by energy-momentum conservation, eq.(\ref{k}). It is
also apparent from the previous equation the cubic dependence on the photon
three-momentum $|\vec{k}|$.
 
From eq.(\ref{widthR}) we can also define the differential decay widths,
\be
\label{ddw}
\frac{d\Gamma(\phi \rightarrow \gamma R)}{d|\vk|}=\frac{M_\phi}{\sqrt{Q^2}} 
\frac{d\Gamma(\phi \rightarrow \gamma R)}{d \sqrt{Q^2}}=
\frac{|\vk| d|\vk|}{24\pi^2M_\phi}\hbox{Imag}\,[t_{22}^R(Q^2)] 
\left|\cH^I(Q^2)\right|^2(M_\phi^2-Q^2)^2~.
\ee
Note that $\hbox{Imag}\,t_{22}^R$ extends from the two pion threshold for 
the $f_0(980)$ case and from the $\pi\eta$ one for the $a_0(980)$ resonance.

As stated above, the standard two body decay formula for $\phi\rightarrow \gamma R$,
with a well defined mass $m_R$ for $R$, can be obtained from eq.(\ref{widthR}) by
performing the substitution $f_R(Q^0)=\delta(Q^0-w_R)$. In this case the decay
width is given by:
\be
\label{gama2body}
\Gamma(\phi\rightarrow \gamma R)_{m_R\,fixed}=
\frac{|\vk|^3}{12\pi}\left|\cH^I(m_R^2)\,
g_{K^+ K^-}^R\right|^2~.
\ee


Following the previous lines, we can also derive an expression for the decay
widths $\Gamma(\phi\rightarrow \gamma M^0N^0)$  by considering {\it
  explicitly} that the strong
amplitudes $t^R_{K^+K^-\rightarrow M^0N^0}$ are dominated by the exchange of
the corresponding $R$ resonances. One should keep in mind that the results
given in sec.\ref{sec:direct} do not make such assumption and are derived {\it 
directly} in terms of the corresponding strong amplitudes
$t^R_{K^+K^-\rightarrow M^0N^0}$. 

Let us denote by $\vq_1$ and $\vq_2$ the three-momenta of the $M^0$ and $N^0$
particles and by $w_1$ and $w_2$ their corresponding energies. Then, since we 
are considering the decay $\phi \rightarrow \gamma M^0 N^0$ to be mediated 
by the resonance $R$,  if follows that,
\ba
\label{mediation}
\Gamma(\phi\rightarrow \gamma M^0 N^0)&=&\frac{1}{3}\sum_{\nu=1}^3 
\sum_{\lambda=1}^2 \frac{1}{2M_\phi}\int \frac{d\vk}{(2\pi)^3 2|\vk|} 
\left|\frac{\cM(\phi\rightarrow \gamma R)}{\cD_R(Q^2)} \right|^2 
\,2 \sqrt{Q^2} \nn \\
&\times&\left\{ \frac{1}{2 \sqrt{Q^2}} \int \frac{d\vp_1}{(2\pi)^3 2w_1}
  \frac{2\pi \delta(Q^0-w_1-w_2)}{2w_2} |g_{M^0 N^0}^R|^2 \right\}~,
\ea
where $\cD_R(Q^2)$ is any possible inverse propagator for the resonance $R$. Let us note that 
the term between curly brackets just corresponds to the standard formula for the decay
width of a resonance of mass $\sqrt{Q^2}$ to the $M^0 N^0$ state with a coupling 
$g_{M^0 N^0}^R$. In addition, we can rewrite, 
\ba
\label{new}
\left|\frac{\cM}{\cD_R(Q^2)}\right|^2&=&
\left|\frac{\cM'\, g_{K^+K^-}^R(Q^2)}{\cD_R(Q^2)}\right|^2=
\left| \cM' \right|^2 \,\left|t_{22}^R(Q^2)\right|^2~. 
\ea
On the other hand, within our present assumption that $t_{22}^R(Q^2)$ is dominated by the exchange 
of the scalar resonance $R$, unitarity implies:
\ba
\label{propa}
\hbox{Imag}[t_{22}^R(Q^2)]&=&\frac{1}{2}\sum_\alpha\, t_{2\alpha}^R(Q^2)^* 
\frac{q_\alpha\, \theta(Q^2-W^2_\alpha)}{4 \pi \sqrt{Q^2}}\, t^R_{\alpha 2}(Q^2)=
\left| t_{22}^R(Q^2)\right|^2 \sum_\alpha 
|g^R_\alpha|^2 \frac{q_\alpha \,\theta(Q^2-W^2_\alpha)}{8\pi \sqrt{Q^2}} \nn \\
&=&|t_{22}^R|^2 \,\sqrt{Q^2} \,\Gamma_{R,tot}(Q^2)~.
\ea
Where $q_\alpha$ is the center of mass three-momentum of the two-body channel $\alpha$, 
$W_\alpha$ is the corresponding threshold energy and $\Gamma_{R\, tot}(Q^2)$ is the total
energy of a scalar resonance of mass $\sqrt{Q^2}$ with the same couplings, 
indicated by $g_\alpha^R$, as the $R$ resonance.

As a result of eqs.(\ref{new}), (\ref{propa}) we can then write instead of eq.(\ref{mediation}):
\ba
\label{propaf}
\Gamma(\phi\rightarrow \gamma M^0N^0)&=&\frac{1}{3}\sum_{\nu=1}^3 
\sum_{\lambda=1}^2\frac{1}{2M_\phi}\int\frac{d \vk}{(2\pi)^3|\vk|}
|\cM'|^2\, \hbox{Imag}\left[t_{22}^R(Q^2)\right] \,
\frac{\Gamma_{R, M^0 N^0}(Q^2)}{\Gamma_{R,tot}(Q^2)}\nn\\
&=&\int 
\frac{|\vk|d|\vk|}{24 \pi^2 M_\phi}|\cH^I(Q^2)|^2\,\hbox{Imag}\left[t_{22}^R(Q^2) \right]
(M_\phi^2-Q^2)^2\,Br_{M^0 N^0}^R(Q^2)~,
\ea
where $\Gamma_{R,M^0 N^0}(Q^2)$ is the decay width of a scalar resonance with mass $\sqrt{Q^2}$ and
the same coupling $g^R_{M^0 N^0}(Q^2)$ to the $M^0 N^0$ channel as the $R$
resonance at that energy. This parameterization can be seen as an alternative to that of
ref.\cite{acha} and is specially suited for those approaches that directly
work with the strong amplitudes, without considering specific forms for the
propagators of the scalar resonances which are explicitly needed in the
legitimate parameterization of ref.\cite{acha}.

Consistently with eq.(\ref{propa}), the branching ratio $Br_{M^0 N^0}^R(Q^2)$ of the strong
decay $R(Q^2)\rightarrow M^0 N^0$ is calculated as:
\be
\label{br}
Br_{M^0 N^0}^R(Q^2)=\frac{|g_{M^0 N^0}^R|^2\, p_{M^0 N^0}}{\sum_\alpha |g_\alpha^R|^2 
p_\alpha \,\theta(Q^2-W_\alpha^2)}~,
\ee
where in addition one should include, when necessary, the corresponding $1/2$
factors for the decays to symmetric two particle states. For the sake of
completeness we show in table \ref{tab:coup}  the pole positions, 
$\sqrt{s_R}$ and couplings, 
$g_\alpha^R$, for the
$f_0(980)$ and $a_0(980)$ resonances from the T-matrices of
ref.\cite{npa}.\footnote{Since the $I=0$ $\pi\pi$ channel is
completely symmetric under the exchange of the pions, the convention of
including an extra factor $1/\sqrt{2}$ in the definition of this state is
followed in ref.\cite{npa}. Thus, the $\pi\pi$ coupling resulting by applying directly
eq.(\ref{residue}) to the $I=0$ T-matrix of ref.\cite{npa} has been
multiplied by $\sqrt{2}$ in table \ref{tab:coup}.}  The latter are defined by:
\be
\label{residue}
|g_{\alpha}^R g_{\beta}^R|=\lim_{s\rightarrow s_R}|(s-s_R)\, t_{\alpha \rightarrow \beta}^R(s)|~,
\ee
where $s_R$ is the pole position of the $R$ resonance in the unphysical sheet
where the center of mass three-momentum of the $M^0 N^0$ state has negative
imaginary part, while that of the $K\bar{K}$ three-momentum is
positive. It is also worth noticing that the couplings $|g^R_{K \bar{K}}|$ are different for the
$f_0(980)$ and $a_0(980)$ resonances like in ref.\cite{referee}. 

Taking from table \ref{tab:coup} the values of the couplings and pole positions
one calculates from eq.(\ref{br}), $Br^{f_0}_{\pi\pi}=0.7$ and $Br^{a_0}_{\pi\eta}=0.63$ in
perfect agreement with the branching ratios given in ref.\cite{npa} where 
the finite widths of the $f_0(980)$ and $a_0(980)$ are taken into
account for the strong decay rates as well.

\begin{table}[H]
\begin{center}
\begin{tabular}{|c|c|}
\hline
$f_0(980)$ & $a_0(980)$ \\
\hline
$\sqrt{s_{f_0}}=(992.6-i\,11.8)$  MeV & $\sqrt{s_{a_0}}=(1009.2-i\,56.1)$
MeV\\
$|g_{\pi\pi}^{f_0}|=1.90$ GeV  & $
|g_{\pi\eta}^{a_0}|=3.54 \hbox{ GeV}$ \\
$|g_{K\bar{K}}^{f_0}|=3.80 \hbox{ GeV}$  &$|g_{K\bar{K}}^{a_0}|=5.20 \hbox{ GeV}$\\
\hline   
\end{tabular}
\caption{Poles positions, $\sqrt{s_R}$, and couplings, $g_{\alpha}^R$, of
  the $f_0(980)$ and $a_0(980)$ resonances from the T-matrices of ref.\cite{npa}.
\label{tab:coup}  }
\end{center}
\end{table}

Comparing eq.(\ref{propaf}) with eq.(\ref{widthR}) for the decay $\phi \rightarrow \gamma R$ 
one obtains the relation,
\be
\label{rel}
\frac{d\Gamma(\phi\rightarrow \gamma M^0
  N^0)}{d|\vk|}=\frac{d\Gamma(\phi\rightarrow \gamma R)}{d|\vk|}\, Br^R_{M^0 N^0}(Q^2)~,
\ee
which, together with eq.(\ref{ddw}), can  be used to analyze in a pure phenomenological way 
the experimental  data of $\phi\rightarrow \gamma M^0 N^0$ by parameterizing and fitting
$\hbox{Imag}[t_{K^+ K^- \rightarrow K^+ K^-}^R]$, as e.g. in
  eq.(\ref{achaprop}). Because we are assuming that the decay $\phi \rightarrow \gamma M^0 N^0$ is
  dominated by the exchange of the corresponding $R$ resonance,  the previous relation
  tells us that the differential width $\phi\rightarrow \gamma
  M^0 N^0$ is the same as the
  one to $\gamma R$ multiplied by the probability that the scalar resonance decays to
$M^0 N^0$. The latter is just the branching ratio since the resonance
  will certainly decay to any channel at asymptotic times. This natural result
  is indeed the way the experimental values on the $\phi\rightarrow \gamma
  f_0(980)$ and $\phi\rightarrow \gamma a_0(980)$ widths are obtained once the
  rates $\phi\rightarrow \gamma f_0(980)\rightarrow \gamma \pi^0\pi^0$ and
$\phi\rightarrow \gamma a_0(980)\rightarrow \gamma  \pi^0\eta$, respectively,
  are determined by fitting the experimental data
  \cite{akh,exppi0,expeta,dafnea0}.\footnote{In ref.\cite{dafnef0} an extra
  and large $\sigma \gamma$
  contribution is included in addition to the $f_0(980) \gamma$ one with a
  destructive interference at low $\pi\pi$ invariant mass. However,
  this is a controversial result at odds with the conclusions of
the previous experimental analysis  refs.\cite{exppi0,snd98}, giving 
rise to an unconventionally large rate $\phi \rightarrow \gamma f_0 (980)$
  \cite{close}, larger than the ones given from any model.} 
Nevertheless, one should keep in mind that eq.(\ref{rel}), when used with the
  simplified form of $Br^R_{M^0N^0}$ given in eq.(\ref{br}), is an
  approximation since it does not take into 
account the finite widths of the $f_0(980)$ and $a_0(980)$ resonances in their 
strong decay rates.

Let us stress that eq.(\ref{gammam0n0}) is given in terms of the strong 
S-wave T-matrices of ref.\cite{npa,iamprd} without any reference to the exchange of an 
scalar resonance $R$. Indeed, in ref.\cite{npa} the scalar resonances
$f_0(980)$ and $a_0(980)$, together with the $\sigma$ meson, are generated dynamically 
as meson-meson resonances in terms of an interacting kernel given by the lowest order 
CHPT amplitudes \cite{wein,gl} without any explicit resonance field. 
Thus, it is an important consistency check whether the invariant mass
distribution obtained from eq.(\ref{gammam0n0}) agrees indeed
with the one of eq.(\ref{rel}) when substituting in the latter 
$d\Gamma(\phi \rightarrow \gamma R)/d|\vk|$ from eq.(\ref{ddw}). 
It is straightforward to demonstrate by  relating 
Imag$[t^R_{K^+K^-\rightarrow K^+ K^-}]$ and $|t^R_{K^+K^-\rightarrow M^0
  N^0}(Q^2)|^2$ via unitarity, 
that eq.(\ref{gammam0n0}) and the combination of eqs.(\ref{rel}) and (\ref{ddw}) yield 
the same $d\Gamma(\phi \rightarrow \gamma M^0N^0)/d|\vk|$ below the $K\bar{K}$ 
threshold. Above it, a specific energy dependent form of $Br^R_{M^0 N^0}$ should be employed.

\begin{figure}[htb]
\psfrag{gamaf0}{\hspace{-1.8cm}{\small $d Br(\phi\rightarrow \gamma \pi^0\pi^0)/dm_{\pi\pi}
\,10^8$}}
\psfrag{gamaa0}{\hspace{-1.8cm}{\small $d Br(\phi\rightarrow \gamma \pi^0\eta)/dm_{\pi\eta}
\, 10^8$}}
\psfrag{E(MeV)}{\begin{tabular}{l} \\ \hspace{-0.5cm}{\small $m_{\pi\pi}$ (MeV)}\end{tabular}}
\psfrag{Y}{\begin{tabular}{l} \\ \hspace{-0.5cm}{\small $m_{\pi\eta}$ (MeV)}\end{tabular}}
\centerline{\epsfig{file=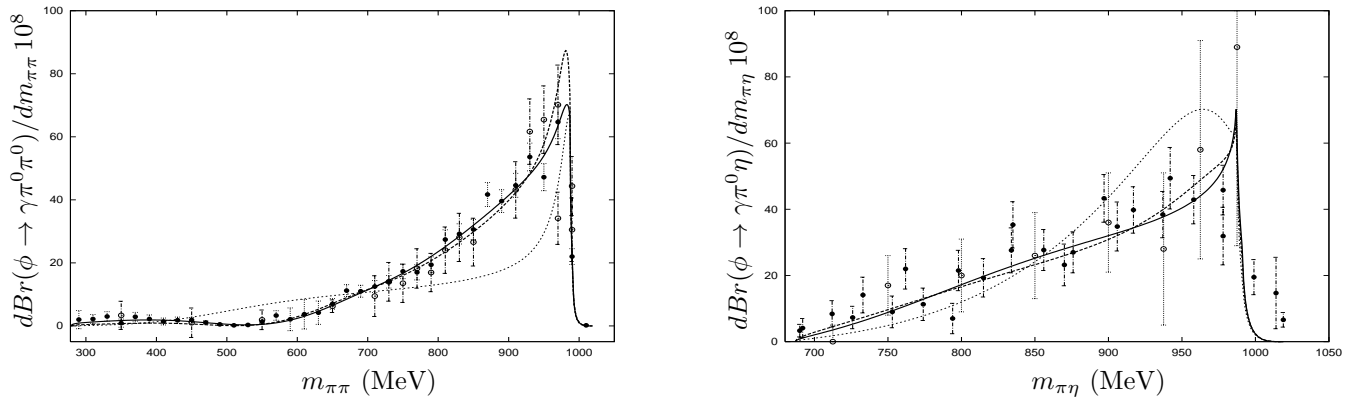,width=7.0in}}
\vspace{.3cm}
\caption[pilf]{\protect \small Invariant mass distributions $d Br(\phi\rightarrow
\gamma \pi^0\pi^0)/dm_{\pi\pi}\, 10^8$ and $d Br(\phi\rightarrow \gamma
\pi^0\eta)/d m_{\pi\eta}\,10^8$
from left to right, respectively. The thick lines correspond to a simultaneous fit of the
parameters $\zeta_I$ and $\delta G^I$ to the experimental points
of \cite{exppi0} (empty circles) and \cite{dafnef0} (full circles), 
for the $\gamma \pi^0\pi^0$ final state,
and to the data of \cite{expeta} (empty circles) and \cite{dafnea0} (full
circles) for the $\gamma \pi^0\eta$ one.
 The thin line corresponds to $\zeta_I=-G_V/\sqrt{8}$ and $\delta G^I=0$,
the appropriate values to reproduce the results of ref.\cite{marco}.
\label{fig:fit}}
\end{figure}


\section{Results and discussion}
\label{sec:wfr}
\setcounter{equation}{0}

In sec.\ref{sec:direct} we have derived the corresponding expressions for the 
$\phi\rightarrow \gamma M^0 N^0$ decay by taking into account the final state
interactions due to the strong S-wave meson-meson amplitudes. This is what we
denote by the scalar contribution to such decay rates. Nevertheless, for
the $\gamma \pi^0\pi^0$ final state the background $\phi\rightarrow \pi^0 \rho
\rightarrow \gamma \pi^0\pi^0$ is not negligible, its relative size depending 
strongly on the energy region, as we show below in
fig.\ref{fig:fitconstraint}. In our final results we have included the
interference term between the scalar and $\rho\pi^0$ contributions with our
own scalar amplitudes and with the vector part calculated as in
ref.\cite{escri}. For the appropriate formulae for the vector piece we refer
to that reference.

\begin{figure}[htb]
\psfrag{distributionf0}{\hspace{-1.6cm}{\small $d Br(\phi\rightarrow \gamma f_0)/d\sqrt{Q^2}
\,10^8$}}
\psfrag{distributiona0}{\hspace{-1.6cm}{\small $d Br(\phi\rightarrow \gamma a_0)/d\sqrt{Q^2}
\, 10^8$}}
\psfrag{E(MeV)}{\begin{tabular}{l} \\ \hspace{-0.5cm}{\tiny $\sqrt{Q^2}$ (MeV)}\end{tabular}}
\centerline{\epsfig{file=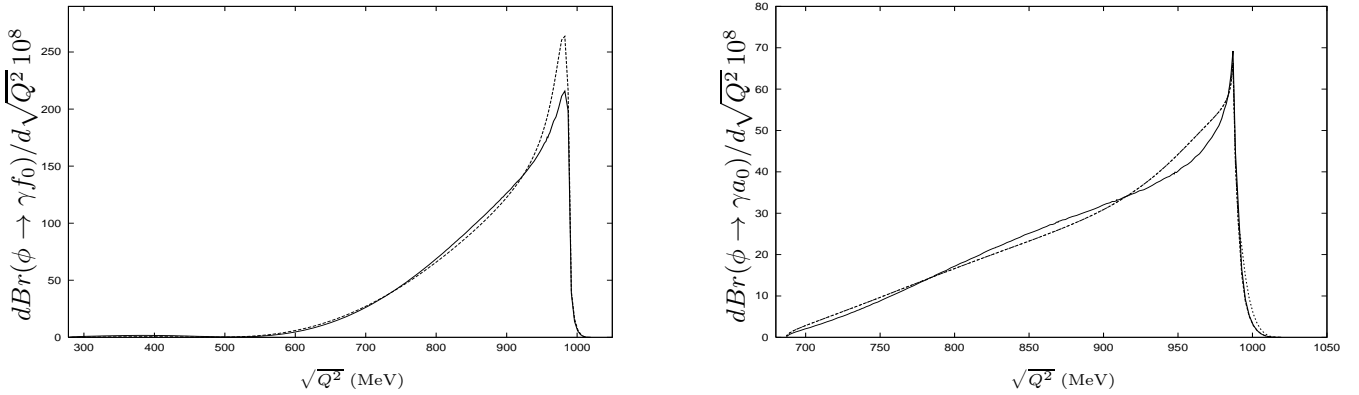,width=7.0in}}
\vspace{.3cm}
\caption[pilf]{Invariant mass distributions $d Br(\phi\rightarrow
\gamma f_0(980))/d\sqrt{Q^2}\, 10^8$ and $d Br(\phi\rightarrow \gamma a_0(980))/d\sqrt{Q^2}\,10^8$
from left to right, respectively. The solid and dashed lines are calculated from
the T-matrices of the refs.\cite{iamprd,npa}, respectively.
\label{fig:disR}}
\end{figure}

We now perform a simultaneous fit, including both the scalar and vector
contributions, to the $\phi\rightarrow \gamma \pi^0\pi^0$ data of 
refs.\cite{exppi0,dafnef0} and those of $\phi\rightarrow
\gamma\pi^0\eta$ of refs.\cite{expeta,dafnea0}. The fit to the invariant mass
distributions $d\Gamma(\phi\rightarrow\gamma\pi^0\pi^0)/dm_{\pi\pi}$ and
$d\Gamma(\phi\rightarrow\gamma\pi^0\eta)/dm_{\pi\eta}$ with the T-matrices of
refs.\cite{npa,iamprd} is presented in fig.\ref{fig:disR} in the left and right
panel, respectively. The solid lines corresponds to the final state
interactions calculated from the scalar amplitudes of
ref.\cite{iamprd} and the dashed ones to the amplitudes of ref.\cite{npa}. The fit reproduces
fairly well the experimental invariant mass distributions and the resulting
values of the parameters are:

\begin{table}[H]
\begin{center}
\begin{tabular}{cll}
T-matrix & $ \zeta_0=+164.12\hbox{ MeV}$  & $\delta G^0=1.46/(4 \pi)^2$  \\
ref.\cite{npa} & $\zeta_1=-165.87\hbox{ MeV}$  & $\delta G^1=1.36/(4 \pi)^2$ \\
& & \\
T-matrix & $\zeta_0=+124.99\hbox{ MeV}$ & $\delta G^0=1.61/(4 \pi)^2$ \\
ref.\cite{iamprd} & $\zeta_1=-132.26\hbox{ MeV}$ & $\delta G^1=1.44/(4 \pi)^2$ \\
\end{tabular}
\caption{Values of the parameters from a simultaneous fit to
  $d\Gamma(\phi\rightarrow\gamma\pi^0\pi^0)/dm_{\pi\pi}$
\cite{exppi0,dafnef0} and
  $d\Gamma(\phi\rightarrow\gamma\pi^0\eta)/dm_{\pi\eta}$
  \cite{expeta,dafnea0}.
\label{tab:fits}}
\end{center}
\end{table}
From the previous table of values is clear an obvious symmetry in the
results. We have obtained that $\delta G^0\simeq \delta G^1$ and
$\zeta_0\simeq -\zeta_1$. The $\zeta_I$ parameters represent the direct
coupling $\phi\gamma K\bar{K}$ with isospin $I$, eq.(\ref{local}). Hence the
direct coupling for the $K^+K^-$ channel as given by eq.(\ref{local}) and the
values in table \ref{tab:fits} is proportional to
$-(\zeta_1+\zeta_0)/\sqrt{2}\simeq 0$ and vanishes. On the contrary, the one
for $K^0\bar{K}^0$ is $(\zeta_1-\zeta_0)/\sqrt{2}\simeq \sqrt{2}
\zeta_1$. Let us note as well that because $\delta G^0\simeq \delta G^1$ one
has that ${\cal   G}^1\simeq {\cal G}^0$. Since the Clebsch-Gordan
coefficients for $K^+K^-$ are the same
both for $I=0$ and 1 its contribution to any process is just proportional to
${\cal G}^0\zeta_0+{\cal G}^1\zeta_1\simeq 0$ and cancels, as one should expect
since the direct coupling $\phi\gamma K^+K^-$ on the left vertex of the
diagram of fig.\ref{fig:loop2}
vanishes as we have just seen. Then it is clear the
consistency of having obtained within the isospin formalism that 
${\cal G}^1={\cal G}^0$ once $\zeta_0=-\zeta_1$,  otherwise there would have
been a mismatch between the results
obtained from the physical basis of states and that of isospin. For the $K^0\bar{K}^0$
channels, since the Clebsch-Gordan coefficients change of sign when passing
from $I=0$ to $I=1$, one has the non-vanishing result ${\cal G}^1\zeta_1-{\cal
  G}^0\zeta_0\simeq 2 {\cal G}^1\zeta_1$. Thus our results suggest the
existence of a  $\phi\gamma K^0\bar{K}^0$ local term of the same type as
that of eq.(\ref{local}) but with $\zeta_I$ replaced by
$\zeta_{K^0\bar{K^0}}=\sqrt{2}\zeta_1$ and the absence of such terms for 
$K^+ K¯$. The resulting value for $|\zeta_I|$ is very similar to $F_V$ from
$\rho\rightarrow e^+ e^-$, $F_V=154$ MeV or from $\phi\rightarrow e^+ e^-$,
$F_V=165$ MeV. The value for $\delta G^I$ is of natural size since it is a
number of order one over $16 \pi^2$. It is also important to remark that in
the reproduction of the lowest
energy part of the $d\Gamma(\phi\rightarrow\gamma\pi^0\pi^0)/dm_{\pi\pi}$
invariant mass distribution, namely below 600 MeV, the background $\rho\pi^0$
plays a significant role, while its negligible above 700 MeV in agreement with
ref.\cite{referee}. From the integration of the invariant mass
distributions we obtain the branching ratios:
\ba
\label{brfiiam}
Br(\phi\rightarrow\gamma\pi^0\pi^0)&=& 1.09 \,10^{-4}~,\nn\\
Br(\phi\rightarrow\gamma\pi^0\eta)&=&0.72 \, 10^{-4}~.
\ea
for the T-matrices of ref.\cite{iamprd}. The values of the previous branching ratios when using
the strong amplitudes of ref.\cite{npa} are almost the same:
\ba
\label{brfinpa}
Br(\phi\rightarrow\gamma\pi^0\pi^0)&=& 1.09 \,10^{-4}~,\nn\\
Br(\phi\rightarrow\gamma\pi^0\eta)&=&0.73 \, 10^{-4}~.
\ea
For comparison, the precise experimental results from
refs.\cite{dafnef0,dafnea0} are $Br(\phi\rightarrow\gamma \pi^0\pi^0)=(1.09\pm0.03_{stat}\pm
0.05_{sys})\,10^{-4}$ and $Br(\phi\rightarrow\gamma \pi^0\eta)=(0.796 \pm
0.07)\,10^{-4}$, respectively, in excellent agreement with our results.

In fig.\ref{fig:fitconstraint} we present a new fit to the same data as before
but imposing the constraints $\delta G^0=\delta G^1$ and
$\zeta_0=-\zeta_1$ that  have emerged in a consistent way
from the fit described above. The values obtained are:
\ba
\label{fitconst}
\hbox{T-matrix of ref.\cite{npa}} &\zeta_0=-\zeta_1=+180.83\hbox{ MeV}~, \,\,
\delta G_0=\delta G_1=1.42/(4\pi)^2~,\nn\\
\hbox{T-matrix of ref.\cite{iamprd}}    &\zeta_0=-\zeta_1=+146.42\hbox{ MeV}~, \,\,
\delta G_0=\delta G_1=1.54/(4\pi)^2~.
\ea
 The quality of the fit is similar to that of
fig.\ref{fig:fit}. In the same figure we also show by the thin dotted line the
$\rho\pi^0$ intermediate contribution which as told above turns out to be relevant only
in the lowest energy range of the spectrum. Let us note as well that in
the experimental analyses at least 6 free parameters \cite{akh,exppi0,expeta} are
needed to fit the experimental $\phi\rightarrow\gamma\pi^0\eta$ and
$\gamma\pi^0\pi^0$ invariant mass distributions.

\begin{figure}[htb]
\psfrag{gamaf0}{\hspace{-1.8cm}{\small $d Br(\phi\rightarrow \gamma \pi^0\pi^0)/dm_{\pi\pi}
\,10^8$}}
\psfrag{gamaa0}{\hspace{-1.8cm}{\small $d Br(\phi\rightarrow \gamma \pi^0\eta)/dm_{\pi\eta}
\, 10^8$}}
\psfrag{X}{\begin{tabular}{l} \\ \hspace{-0.5cm}{\small $m_{\pi\pi}$ (MeV)}\end{tabular}}
\psfrag{Y}{\begin{tabular}{l} \\ \hspace{-0.5cm}{\small $m_{\pi\eta}$ (MeV)}\end{tabular}}
\centerline{\epsfig{file=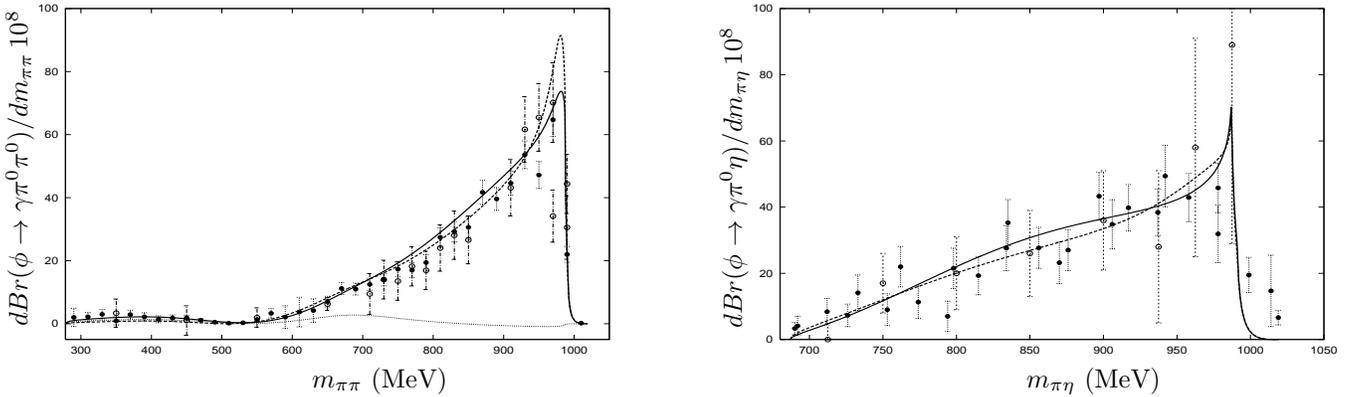,width=7.0in}}
\vspace{.3cm}
\caption[pilf]{\protect \small Invariant mass distributions $d Br(\phi\rightarrow
\gamma \pi^0\pi^0)/dm_{\pi\pi}\, 10^8$ and $d Br(\phi\rightarrow \gamma
\pi^0\eta)/d m_{\pi\eta}\,10^8$
from left to right, respectively. The thick lines correspond to a simultaneous fit of the
parameters $\zeta_I$ and $\delta G^I$ to the experimental points
of \cite{exppi0} (empty circles) and \cite{dafnef0} (full circles), for the
$\gamma \pi^0\pi^0$ final state,
and to the data of \cite{expeta} (empty circles) and \cite{dafnea0} (full circles) for
 the $\gamma \pi^0\eta$ one. The constraints $\delta G^1=\delta G^0$ and 
$\zeta_0=-\zeta_1$
have been imposed. The thin dotted line corresponds to the $\rho \pi^0$ background.
\label{fig:fitconstraint}}
\end{figure}

Once the $\zeta_I$ and $\delta G^I$  are fixed, table \ref{tab:fits} and eq.(\ref{fitconst}),
we show in fig.\ref{fig:disR}, from left to right,
the distributions $d Br(\phi\rightarrow \gamma f_0(980))/dm \,10^8$ and
$d Br(\phi\rightarrow \gamma a_0(980))/dm \,10^8$, respectively, calculated from
eq.(\ref{ddw}) and the strong amplitudes of ref.\cite{npa,iamprd}. As in figs.\ref{fig:fit}
the solid lines corresponds to ref.\cite{iamprd} and the dashed ones to
ref.\cite{npa}. 
We have denoted by  $m=\sqrt{Q^2}$  the invariant mass of the $f_0(980)$ and
$a_0(980)$ resonances.   It can be
surprising that the tails of the invariant mass distributions
extend well below the prominent peaks of the $f_0(980)$ and $a_0(980)$ resonances towards
rather low invariant masses.
This is due to the cubic dependence on the photon
three-momentum $|\vk|=(M_\pi^2-Q^2)/2M_\phi$ which largely enhances the low
energy part of the invariant mass distributions. Indeed, if we fix $|\vk|$ to the
value corresponding to some  nominal mass of the resonances, e.g. $m_R\simeq 986$ MeV,
then one obtains very peaked distributions around the masses of the $f_0(980)$ and
$a_0(980)$ without any tail towards low energies. Hence, it follows that although
the resonance structure is clear and very prominent the low energy
components of the energy distributions $f_R(Q^0)$, see fig.\ref{fig:f_0},
cannot be neglected because of the enhancement due to the cubic
dependence on the photon three-momentum, fig.\ref{fig:disR}.

We integrate now the invariant mass distributions $d\Gamma(\phi\rightarrow
\gamma R)/d \sqrt{Q^2}$ from the corresponding thresholds
up to $Q^2=M_\phi^2$, eq.(\ref{figamaR}), and we obtain the values shown in
table \ref{table:widths}.
\begin{table}[H]
\begin{center}
\begin{tabular}{|l|c|c|c|}
\hline
 & T-matrix ref.\cite{npa} &  T-matrix ref.\cite{iamprd}\\
\hline
$Br(\phi\rightarrow \gamma f_0(980))\, 10^4$& 3.19  & 3.11  \\
\hline
$ Br(\phi\rightarrow \gamma a_0(980))\, 10^4$& 0.73 &  0.73 \\
\hline
$\frac{ Br(\phi\rightarrow \gamma f_0(980))}{ Br(\phi\rightarrow \gamma
  a_0(980))}$ & 4.37  & 4.26 \\
\hline
\end{tabular}
\caption{$Br(\phi\rightarrow \gamma f_0(980))$, $Br(\phi\rightarrow \gamma
  a_0(980)$ and the quotient of both from eq.(\ref{figamaR}) and the energy
  distribution $f_R(Q^0)$, eq.(\ref{uni2}), determined with the
  T-matrices of refs.\cite{iamprd,npa}.
\label{table:widths}}
\end{center}
\end{table}
In this table the energy distributions $f_R(Q^0)$, eq.(\ref{uni2}), are determined
from the  T-matrices of refs.\cite{npa,iamprd}.  For $\zeta_I=-G_V/\sqrt{8}$ and
$\delta G^I=0$, which corresponds to the results of ref.\cite{marco}, one has
$Br(\phi\rightarrow \gamma f_0(980))= 2.34 \,10^{-4}$,
$Br(\phi\rightarrow \gamma a_0(980))= 0.86 \,10^{-4}$ and the quotient between
both rates is then 2.72.

The SND collaboration, refs.\cite{exppi0} and \cite{expeta}, reports the branching ratios 
$ Br(\phi\rightarrow \gamma f_0(980))= (3.5\pm 0.3 ^{+1.3}_{-0.5}) \,10^{-4}$
and 
$ Br(\phi\rightarrow \gamma a_0(980))= (0.88\pm 0.17) \, 10^{-4} $, 
respectively. The CMD-2 collaboration, ref.\cite{akh}, reports 
$ Br(\phi\rightarrow \gamma f_0(980))= (2.90\pm0.21\pm 1.54)\,10^{-4}$.  
Taking into account the latter value for $Br(\phi \rightarrow \gamma f_0(980))$, as in 
ref.\cite{pdg} since it arises from a combined fit to the
$\gamma \pi^0\pi^0$ and $\gamma \pi^+\pi^-$ data\footnote{In addition,
the error in the result
$Br(\phi\rightarrow \gamma f_0(980))= (2.90\pm0.21\pm 1.54)\,10^{-4}$ of
ref.\cite{akh} is enlarged in this reference so that the rate 
$\phi \rightarrow \gamma f_0$, determined by
performing a narrow pole fit, is also compatible within errors.},
and the one of ref.\cite{expeta} for $ Br(\phi\rightarrow \gamma a_0(980))$,
one has
$Br(\phi\rightarrow \gamma f_0(980))/Br(\phi\rightarrow a_0(980))\simeq 3.3 \pm 2.0$.
 These numbers are based in a parameterization of the experimental
event distributions of $\pi^0\pi^0$ and $\pi^0\eta$ leaving the masses and
couplings of the $f_0(980)$ and $a_0(980)$ as free parameters.  Once these are
determined by the fitting procedure, the branching ratios to $\gamma R$ are
obtained by dividing the resulting $Br(\phi\gamma \rightarrow \gamma R\rightarrow
\gamma M^0 N^0)$ by the corresponding $Br^R_{M^0N^0}$. 
There is some model dependence on the ``experimental'' numbers
because of the specific forms of the resonance propagators.
Nevertheless, we see remarkable agreement between these experimental results 
and our calculations of table \ref{table:widths}. In ref.\cite{dafnef0} the KLOE collaboration
reports the result $Br(\phi \rightarrow \gamma f_0(980))=
(4.47\pm 0.21 )\,10^{-4}$. This value is obtained by including a very specific
 destructive interference in the low energy region between the
$f_0(980)\gamma$ and a $\sigma \gamma$ contribution, both of them very large
at such energies. The $Br(\phi\rightarrow \gamma a_0(980))$
is much clearer and very well established due to the absence of any significant background
contributions to $\phi\rightarrow \gamma \pi^0\eta$ as established both
theoretical and experimentally,
refs.\cite{referee,bra3,escri,expeta,dafnea0}. 
The KLOE collaboration
\cite{dafnea0} determines the  number $Br(\phi \rightarrow \gamma a_0(980)) =
(0.74\pm 0.07)\,10^{-4}$ from the very precise measured rate
$Br(\phi\rightarrow\gamma \pi^0\eta)=(0.796 \pm 0.07)\,10^{-4}$. 
Our calculations for both rates presented in
eq.(\ref{brfiiam}) and in table \ref{table:widths} are in perfect agreement with these
determinations since we described very accurately the corresponding invariant mass
distribution as noted in figs.\ref{fig:fit} and \ref{fig:fitconstraint}.

\begin{center}
\begin{table}[H]
\begin{tabular}{|l||c|c|c|c|}
\hline
& CMD-2--fit 1 \cite{akh}& CMD-2--fit 2 \cite{akh}& SND--\cite{exppi0} &
SND--\cite{expeta}\\
& $Br(\phi\rightarrow \gamma f_0) \,10^{4}$ & $Br(\phi\rightarrow \gamma
f_0) \,10^{4}$ & $Br(\phi\rightarrow \gamma f_0)\,10^{4}$ &
$Br(\phi\rightarrow \gamma a_0) \,10^{4}$  \\
\hline
reported& $2.90\pm 0.21 \pm 0.65$ & $3.05\pm 0.25 \pm 0.72$ &
$4.6 \pm0.3^{+1.3}_{-0.5}$ &  $0.88 \pm 0.17$ \\
\hline
eq.(\ref{figamaR}), $\zeta_I=0$& 3.21 & 3.51 & 4.8 & 0.96\\
\hline
\end{tabular}
\caption{$Br(\phi\rightarrow \gamma f_0(980))$ and $Br(\phi\rightarrow \gamma
  a_0(980))$ branching ratios where the energy distribution $f_R(Q^0)$ is
  fixed from  eq.(\ref{achaprop}) and then eq.(\ref{figamaR}) is used. 
The results reported in the  experimental references \cite{exppi0,expeta}  are
also given.  In the second row of numbers, the function $\cH^I(Q^2)$,
eq.(\ref{figamaR}) is used with $\zeta_I=0$, the choice that corresponds to
 refs.\cite{exppi0,expeta}.
\label{tab:acha}}
\end{table}
\end{center}

 We now apply eq.(\ref{figamaR}) to
the energy distributions $f_R(Q^0)$ coming from eq.(\ref{achaprop}), where the
free parameters, the masses of the $f_0(980)$, $a_0(980)$ and four couplings,
take the values of several fits of refs.\cite{akh,exppi0,expeta} used
to obtain their experimental numbers of the rates $\phi\rightarrow
\gamma R$. The calculation is  performed  with the function
$\cH^I(Q^2)$ calculated as
in refs.\cite{acha,snd98,exppi0,expeta}, that is, with $\zeta_I=0$ in eq.(\ref{figamaR}).
This corresponds to the second row of numerical results of
table \ref{tab:acha}. In this way,
the previous decay widths are calculated without the shortcut to use a
 fixed
value for the branching ratio of the resonance $R$ to the lightest decay
channel  in all the
energy interval up to $M_\phi$, as in the present experimental analyses
 \cite{akh,exppi0,expeta,dafnef0,dafnea0}.

In ref.\cite{exppi0} the quoted $Br(\phi\rightarrow \gamma f_0(980))=(3.5
\pm0.3^{+1.3}_{-0.5})\,10^{-4}$ comes by multiplying by three the measured 
$Br(\phi\rightarrow \gamma
\pi^0\pi^0)$ with $Br^{f_0}_{\pi^0\pi^0}(Q^2)=1/3$, see
eq.(\ref{rel}). In table \ref{tab:acha} the quoted value from
ref.\cite{exppi0} is the one obtained with a fit to $\phi\rightarrow \gamma
\pi^0\pi^0$ invariant
mass distribution including as well a background from $\rho\pi^0$, which amounts
at most to $15\%$. The error is correspondingly
enlarged so as to make both
results compatible. For the $a_0(980)$ case the only number that is reported,
both in table \ref{tab:acha} and ref.\cite{expeta}, is the same as the measured
$Br(\phi\rightarrow \gamma \pi^0\eta)$ assuming $Br^{a_0}_{\pi\eta}(Q^2)=1$. The fit
1 for the CMD-2 collaboration includes both the $\gamma \pi^0\pi^0$ and $\gamma
\pi^+\pi^-$ final states. The CMD-2 second fit includes only the $\gamma
\pi^0\pi^0$ final state. We see a general agreement between the reported and
calculated numbers from eqs.(\ref{achaprop}), (\ref{widthR}) and
(\ref{figamaR}), given in the first and
second rows of numerical results, respectively, although the
latter tend to be somewhat larger, particularly for the reported values of the
CMD-2 collaboration.

It has been recently claimed in ref.\cite{close2} that in order to
interpret the experimental results of refs.\cite{exppi0,expeta} for the rates
$\phi \rightarrow \gamma R$ one needs to include sizeable
isospin violating effects in the couplings of the $f_0(980)$ and $a_0(980)$
resonances to the $K^+ K^-$ and $K^0\bar{K}^0$ channels, so that 
$|g^R_{K^+K^-}|\neq|g^R_{K^0\bar{K}^0}|$ and the difference is claimed to be
as large as a 30$\% $.
Particular emphasis is given to the necessity to deviate from
isospin symmetry in order to understand the quotient between the branching
ratios of the $\phi$ to $\gamma f_0(980)$ and $\gamma a_0(980)$ that, although with a
large experimental uncertainty, as shown above, has a central value of
around three instead of one.
The much more precise results of refs.\cite{dafnef0,dafnea0}, when taking
$Br(\phi\rightarrow\gamma f_0(980))=
3\,Br(\phi\rightarrow \gamma \pi^0\pi^0)$ and
$Br(\phi\rightarrow\gamma a_0(980))=Br(\phi\rightarrow\gamma\pi^0\eta)$, 
imply a value $4.1\pm 0.2$ for this ratio, very close to our results of table 
\ref{table:widths}.
Indeed in our approach the calculated $Br(\phi \rightarrow\gamma f_0(980)$ from
eq.(\ref{figamaR}) and the one obtained by multiplying by three the 
$Br(\phi\rightarrow\gamma\pi^0\pi^0)$
differ in less than a $6\%$. Our study clearly shows that one
can achieve a good agreement with the experimental data without any
deviation from isospin symmetry in the couplings of the $f_0(980)$ and $a_0(980)$ 
resonances to the $K^+ K^-$ and $K^0\bar{K}^0$ channels, although a
contact interaction term $\phi\gamma K^0\bar{K}^0$, beyond the
pure $K^+ K^-$ loop model of ref.\cite{acha}, has to be included as described above. 
It is also clear that one should 
abandon in the study of the $\phi\rightarrow \gamma R$ decays 
the standard two body decay formula, eq.(\ref{gama2body}), used in refs.\cite{close2,joe}, 
 due to the proximity of the threshold of the final 
state to $M_\phi$ and the cubic dependence on $|\vk|$. As a result, 
the effects of the finite
widths of the scalar resonances, and their associated energy distributions $f_R(Q^0)$, 
must be included from the very beginning. 
For instance, had we used the values for the $f_0(980)$ and $a_0(980)$ masses 
and the $g_{K^+K^-}^R$ couplings given in table \ref{tab:coup} we would have
obtained $Br(\phi\rightarrow \gamma f_0)=1.17\, 10^{-4}$ and 
$Br(\phi\rightarrow \gamma a_0)=0.58\, 10^{-5}$. The latter value is so small
due to the rather high pole mass of the $a_0(980)$. Let us stress that this pole mass is 
clearly different to the 
value where the S-wave I=1 T-matrix elements peak, around 986 MeV, which indeed 
changes from one matrix element to the other. Hence, it is rather artificial to decide which 
is the value of $m_R$ to be used for the calculation of the rate 
$\phi \rightarrow \gamma R$ in an extraordinary  sensitive two-body standard decay formula to
the chosen $m_R$ value.

It is worth mentioning that we have reproduced the numerical results of
refs.\cite{plb}, $\zeta_I=0$, and those of ref.\cite{marco}, $\zeta_I=-G_V/\sqrt{8}$,
$\delta G^I=0$. Nevertheless, the $I(a,b)$
function in ref.\cite{plb} was evaluated with the mass of the $K^0$ although
 the mass of the $K^+$ should have been used since it
corresponds to a $K^+K^-$ loop. This kinematical source of isospin violation gives rise
to non-negligible corrections due to the proximity of the $K^0\bar{K}^0$ threshold
to the mass of the $\phi(1020)$. When this is taken into account, one has
$Br(\phi\rightarrow \gamma K^0\bar{K}^0)=3.0 \, 10^{-8}$ instead of $5
\,10^{-8}$ as given in ref.\cite{plb} where the
T-matrices of ref.\cite{npa} were used. We now evaluate the previous branching
ratio with our present formalism and with the values of the $\delta G^I$ and
$\zeta_I$ as given in table \ref{tab:fits}. For the strong amplitudes of
ref.\cite{npa} we obtain $Br(\phi\rightarrow\gamma K^0\bar{K}^0)=3.7\,
10^{-8}$ and with the T-matrices of ref.\cite{iamprd} one has 
$Br(\phi\rightarrow\gamma K^0\bar{K}^0)=6.43\,10^{-9}$. The tree level contact interaction 
 $\phi\gamma K^0\bar{K}^0$ and its iteration
through final state interactions interfere destructively and tend to cancel
each other or even they give rise to a negative interference with the pure $K^+K^-$ kaon
loop contribution of ref.\cite{plb}. 


\section{Conclusions}
\label{sec:conc}
\def\theequation{\Alph{section}.\arabic{equation}}
\setcounter{equation}{0}

In this article we have shown how the experimental
results for the decay widths $\Gamma(\phi \rightarrow \gamma f_0(980))$ and
$\Gamma(\phi \rightarrow \gamma a_0(980))$ from refs.\cite{akh,exppi0,expeta}
can be described 
without abandoning isospin symmetry in the calculation of the S-wave strong T-matrix
elements \cite{npa,iamprd}.  Nevertheless, in our final results we have calculated
the integral $I(a,b)$ in terms of the $K^+$ mass instead of the
average isospin mass. This kinematical isospin violating fact amounts to
effects of around
a 10$\%$ in the rate $\phi\rightarrow \gamma f_0(980)$ and around a 20$\%$ in
the width to $\gamma a_0(980)$.
The $\gamma K^0\bar{K}^0$ branching ratio is much more sensitive to these
kinematical effects due to the so much reduced available phase space.
The same situation would have arisen in the decays $\phi\rightarrow \gamma R$ as well
if we had used the standard two body
decay formula eq.(\ref{gama2body}), with well defined masses $m_R$, instead of
having taken
care of the finite width effects of the $f_0(980)$ and $a_0(980)$ resonances.   These
results are quite opposite to what has been claimed in
ref.\cite{close2} regarding the necessity of including large isospin
violating effects in the couplings of the $f_0(980)$ and $a_0(980)$ scalar resonances
to the $K^+ K^-$ and $K^0\bar{K}^0$ channels due to the proximity of the
$K\bar{K}$ threshold
to the nominal masses of the $f_0(980)$ and $a_0(980)$
resonances. 
We have also stressed that one should abandon the standard
two body decay formula, with well defined masses for the $f_0(980)$ and $a_0(980)$
resonances, in the calculation of the rates $\phi \rightarrow \gamma f_0(980)$ and $\phi
\rightarrow \gamma a_0(980)$. Instead, finite energy distributions, $f_R(Q^0)$, have to be
considered from the very beginning because of the dramatic changes in the resulting
decay widths under small changes of the $f_0(980)$ and $a_0(980)$ masses as compared to their
widths or as compared to the difference between the pole masses and the energy of the
peaks in the S-wave $I=0$ and 1 T-matrices.

It is also worth remarking that the formula derived in sec.\ref{sec:gf} for
 calculating the decay
widths $\phi\rightarrow \gamma R$, eq.(\ref{widthR}), together with eq.(\ref{rel}), can
be also applied in the experimental analyses of the rates $\phi\rightarrow \gamma
R$ as an alternative to the rightful one followed in
refs.\cite{akh,exppi0,expeta,dafnef0,dafnea0} from ref.\cite{acha}.  In this
 way, one does not use, even as an intermediate step, the bare theoretical concept 
$\Gamma(\phi\rightarrow \gamma R;\sqrt{Q^2})$ and incorporates the contribution
 to $\cH^I(Q^2)$ proportional to $\zeta_I$ in eq.(\ref{figamaR}).

In order to describe the data we have included, beyond the $K^+K^¯$ loop model
of ref.\cite{acha}, self-gauge invariant vertices  
$V_I(\phi\gamma K\bar{K})$ with the $K\bar{K}$ pair in the isospin channel
$I$.  As a result of the fit to the experimental data of 
refs.\cite{exppi0,dafnef0,expeta,dafnea0}, a non-vanishing $V(\phi\gamma K^0\bar{K}^0)$ local
term has emerged while the corresponding $V(\phi\gamma K^+K^-)$ tend to vanish. It
is also shown that the former plays an important role in order to reproduce
the experimental data. We have also included this contribution to
the $\phi\rightarrow \gamma K^0\bar{K}^0$ decay although its effects do not spoil here the
conclusions of ref.\cite{plb}, where only the $K^+K^-$ loop
contribution is included, that this branching ratio is negligible small and does
not offer any significant background to study CP violation in a
$\phi$-factory like DA$\Phi$NE.

It has been repeatedly stated that the study of the $\phi(1020)$ radiative decays to
$\gamma R$ constitutes an important test to unveil the nature of the $f_0(980)$
and $a_0(980)$ resonances by comparing the resulting experimental data with
the models and approaches present in the literature. Indeed, we see from
eq.(\ref{figamaR}) that the study of these decays constitutes an
alternative source of experimental information on the $f_0(980)$  resonance since
 it is sensitive to $\hbox{Imag}[t_{22}^R]$ which is not
directly measured in $\pi\pi$ scattering data. For the $a_0(980)$ resonance the experimental
data is much more scarce than for the $f_0(980)$ and hence having new 
precise data, as that of ref.\cite{dafnea0}, is of foremost importance. 
Our simultaneous study of the new and accurate data of
\cite{dafnea0} on $\phi\rightarrow\gamma\pi^0\eta$ together with that of
$\phi\rightarrow\gamma\pi^0\pi^0$ \cite{exppi0,dafnef0} has given a coherent 
reproduction of these  data in terms of only two free parameters. 
This constitutes a step forward in the experimental
verification of the strong T-matrices of refs.\cite{iamprd,npa} already successfully tested in 
refs.\cite{npa,gamma,jpsi,iamprd}.

\medskip
\noindent {\bf Acknowledgments}

\medskip
I would like to thank E. Oset for a critical reading of the manuscript 
and San Fu Tuan for useful communications. This work has been partially 
supported by the EU TMR network Eurodaphne, contract no. ERBFMRX-CT98-0169.

\end{document}